\documentclass[sigconf]{acmart}

\AtBeginDocument{
  }

\copyrightyear{2018}
\acmYear{2018}
\acmDOI{XXXXXXX.XXXXXXX}

\acmConference[Conference acronym 'XX]{Make sure to enter the correct
  conference title from your rights confirmation emai}{June 03--05,
  2018}{Ithaca, NY}

\acmISBN{978-1-4503-XXXX-X/18/06}

\usepackage{algorithm}
\usepackage{algorithmic}
\usepackage{xcolor}
\usepackage{pifont}
\usepackage[utf8]{inputenc}
\usepackage{natbib}
\usepackage{caption}
\usepackage{graphicx}

\usepackage{booktabs}
\usepackage{multirow}
\usepackage{amsmath} 
\usepackage{amsfonts}
\usepackage{amsthm}
\usepackage{bm}
\usepackage{enumitem}
\usepackage{subcaption}

\usepackage{xcolor}      
\usepackage{framed}      
\usepackage{bm}          
\usepackage{fancybox}

\definecolor{recgreen}{RGB}{34,139,34}      
\definecolor{recbggreen}{RGB}{240,255,240}  
\definecolor{epblue}{RGB}{25,25,112}        
\definecolor{epbgblue}{RGB}{240,248,255}    
\definecolor{ipred}{RGB}{178,34,34}         
\definecolor{ipbgred}{RGB}{255,240,240}     
\definecolor{cfolive}{RGB}{143,151,121}     
\definecolor{cfbgolive}{RGB}{248,248,240}   

\newcommand{\recboxsimple}[2][Recommendation prompt $P_{REC}$]{
    \vspace{8pt}
    \noindent
    \setlength{\fboxsep}{0pt}
    \setlength{\fboxrule}{1pt}
    
    \fcolorbox{recgreen}{recbggreen}{
        \begin{minipage}{\dimexpr\linewidth-2\fboxrule}
            
            \colorbox{recgreen}{
                \begin{minipage}{\linewidth}
                    \vspace{5pt}
                    \hspace{8pt}\textcolor{white}{\textbf{#1}}
                    \vspace{5pt}
                \end{minipage}
            }
            
            \vspace{8pt}
            \hspace{8pt}
            \centering
            \begin{minipage}{\dimexpr\linewidth-16pt}
                \textit{#2}
            \end{minipage}
            \vspace{8pt}
        \end{minipage}
    }
    }

\newcommand{\epboxsimple}[2][EP reflection prompt $P_{EP}$]{
    \vspace{8pt}
    \noindent
    \setlength{\fboxsep}{0pt}
    \setlength{\fboxrule}{1pt}
    \fcolorbox{epblue}{epbgblue}{
        \begin{minipage}{\dimexpr\linewidth-2\fboxrule}
            
            \colorbox{epblue}{
                \begin{minipage}{\linewidth}
                    \vspace{5pt}
                    \hspace{8pt}\textcolor{white}{\textbf{#1}}
                    \vspace{5pt}
                \end{minipage}
            }
            
            \vspace{8pt}
            \hspace{8pt}
            \centering
            \begin{minipage}{\dimexpr\linewidth-16pt}
                \textit{#2}
            \end{minipage}
            \vspace{8pt}
        \end{minipage}
    }
    \vspace{3pt}
}

\newcommand{\ipboxsimple}[2][IP reflection prompt $P_{IP}$]{
\vspace{5pt}
    \noindent
    \setlength{\fboxsep}{0pt}
    \setlength{\fboxrule}{1pt}
    \fcolorbox{ipred}{ipbgred}{
        \begin{minipage}{\dimexpr\linewidth-2\fboxrule}
            
            \colorbox{ipred}{
                \begin{minipage}{\linewidth}
                    \vspace{5pt}
                    \hspace{8pt}\textcolor{white}{\textbf{#1}}
                    \vspace{5pt}
                \end{minipage}
            }
            
            \vspace{5pt}
            \hspace{8pt}
            \centering
            \begin{minipage}{\dimexpr\linewidth-16pt}
                \textit{#2}
            \end{minipage}
            \vspace{5pt}
        \end{minipage}
    }
    \vspace{5pt}
}

\newcommand{\cfboxsimple}[2][CF reflection prompt $P_{CF}$]{
\vspace{5pt}
    \noindent
    \setlength{\fboxsep}{0pt}
    \setlength{\fboxrule}{1pt}
    \fcolorbox{cfolive}{cfbgolive}{
        \begin{minipage}{\dimexpr\linewidth-2\fboxrule}
            
            \colorbox{cfolive}{
                \begin{minipage}{\linewidth}
                    \vspace{5pt}
                    \hspace{8pt}\textcolor{white}{\textbf{#1}}
                    \vspace{5pt}
                \end{minipage}
            }
            
            \vspace{5pt}
            \hspace{8pt}
            \centering
            \begin{minipage}{\dimexpr\linewidth-16pt}
                \textit{#2}
            \end{minipage}
            \vspace{5pt}
        \end{minipage}
    }
    \vspace{2pt}
}

\usepackage{newfloat}
\usepackage{listings}
\DeclareCaptionStyle{ruled}{labelfont=normalfont,labelsep=colon,strut=off} 
\floatstyle{ruled}
\newfloat{listing}{tb}{lst}{}
\floatname{listing}{Listing}

\title{MoRE: A Mixture of Reflectors Framework for Large Language Model-Based
Sequential Recommendation}

\newcounter{sharedfootnote}
\author{Weicong Qin}
\authornote{The first two authors contributed equally to this research.}
\author{Yi Xu}
\authornotemark[1]
\affiliation{
  \department{Gaoling School of Artificial Intelligence}
  \institution{Renmin University of China}
  \city{}
  \country{}
}
\email{qwc@ruc.edu.cn}
\email{yixu00@ruc.edu.cn}

\author{Weijie Yu}
\authornote{The corresponding author is Weijie Yu.}
\affiliation{
\department{School of Information Technology and Management}
  \institution{University of International Business and Economics}
  \city{}
  \country{}}
\email{yu@uibe.edu.cn}

\author{Chenglei Shen}
\affiliation{
    \department{Gaoling School of Artificial Intelligence}
  \institution{Renmin University of China}
  \city{}
  \country{}
}
\email{chengleishen9@ruc.edu.cn}

\author{Xiao Zhang}
\affiliation{
    \department{Gaoling School of Artificial Intelligence}
  \institution{Renmin University of China}
  \city{}
  \country{}
}
\email{zhangx89@ruc.edu.cn}

\author{Ming He}
\author{Jianping Fan}
\affiliation{
  \institution{AI Lab at Lenovo Research}
  \city{}
  \country{}}
\email{heming01@foxmail.com}
\email{jfan1@lenovo.com}

\author{Jun Xu}
\affiliation{
  \department{Gaoling School of Artificial Intelligence}
  \institution{Renmin University of China}
  \city{}
  \country{}
}
\email{junxu@ruc.edu.cn}

\renewcommand{\shortauthors}{Qin, Xu and Yu et al.}

\begin{abstract}
Large language models (LLMs) have emerged as a cutting-edge approach in sequential recommendation, leveraging historical interactions to model dynamic user preferences. Current methods mainly focus on learning processed recommendation data in the form of sequence-to-sequence text. While effective, they exhibit three key limitations: 1) failing to decouple \textit{intra-user} explicit features (e.g., product titles) from implicit behavioral patterns (e.g., brand loyalty) within interaction histories; 2) underutilizing \textit{cross-user} collaborative filtering (CF) signals; and 3) relying on inefficient reflection update strategies. 
To address this, We propose \texttt{MoRE} (\underline{M}ixture \underline{o}f \underline{RE}flectors), which introduces three perspective-aware offline reflection processes to address these gaps. This decomposition directly resolves Challenges 1 (explicit/implicit ambiguity) and 2 (CF underutilization). Furthermore, \texttt{MoRE}'s meta-reflector employs a self-improving strategy and a dynamic selection mechanism (Challenge 3) to adapt to evolving user preferences.
First, two \textit{intra-user} reflectors decouple explicit and implicit patterns from a user's interaction sequence, mimicking traditional recommender systems' ability to distinguish surface-level and latent preferences. A third \textit{cross-user} reflector captures CF signals by analyzing user similarity patterns from multiple users' interactions. To optimize reflection quality, \texttt{MoRE}'s meta-reflector employs a offline self-improving strategy that evaluates reflection impacts through comparisons of presence/absence and iterative refinement of old/new versions, with a online contextual bandit mechanism dynamically selecting the optimal perspective for recommendation for each user.
Experiments on three benchmarks show \texttt{MoRE} outperforms both traditional recommenders and LLM-based methods with minimal computational overhead, validating its effectiveness in bridging LLMs' semantic understanding with multidimensional recommendation principles. Code: \url{https://github.com/E-qin/MoRE-Rec}.

\end{abstract}



\begin{document}

\maketitle

\section{Introduction} \label{sec:intro}

Sequential recommendation (SeqRec), which predicts the next item of interest based on a user's historical interaction sequence, is crucial for recommender systems. The key to this task is to capture users’ dynamic preferences accurately.
Large language models (LLMs) are emerging as promising recommenders (LLMREC) due to their vast world knowledge and excellent reasoning abilities. Two typical approaches in this context are prompt-based methods and fine-tuning methods. Prompt-based methods~\cite{dai2023uncovering,hou2023large} pre-construct fixed prompts and exploit in-context learning~\cite{dong2022survey} to guide the LLM in reasoning the desired item. Fine-tuning methods~\cite{zheng2024adapting,zhang2024text,liao2024llara} inject domain knowledge by fine-tuning an LLM on substantial annotated recommendation data. Although these methods achieve encouraging performance, the former suffers from an inability to optimize the prompts based on user feedback, while the latter requires substantial computational resources.

\begin{figure}
    \centering
\includegraphics[width=.95\linewidth]{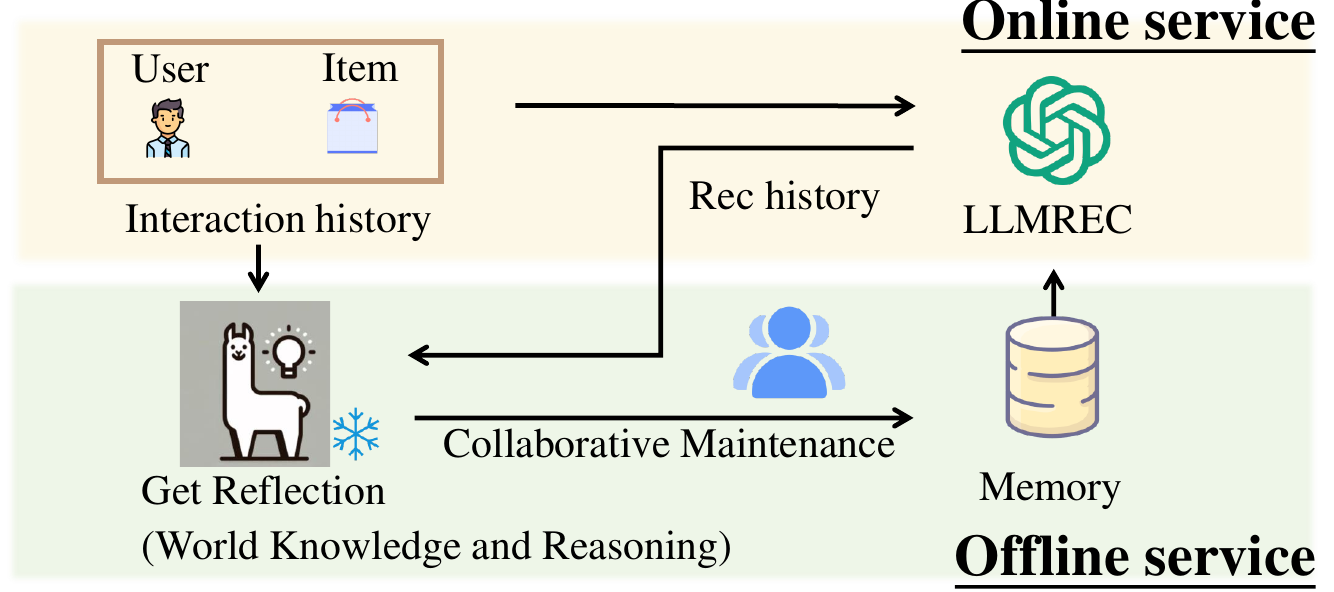}
    \caption{Workflow of reflection-based methods for SeqRec. LLM reflections are generated offline based on interaction and recommendation history. These reflections are then collected as hints to improve online recommendations.}
    \label{fig:more:simple}
\end{figure}

\begin{figure}[t]
    \centering
    \setlength{\belowcaptionskip}{-10pt}
\includegraphics[width=1.0\linewidth]{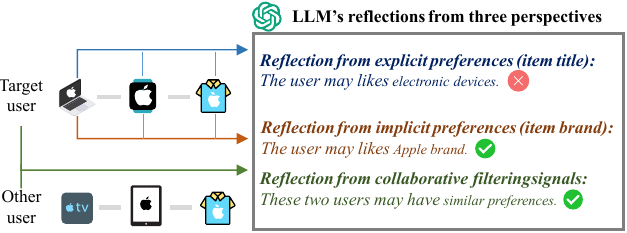}
    \caption{Examples of LLM reflections from explicit, implicit, and CF perspectives based on user interactions.}
    \label{fig:more:bc}
\end{figure}

Recently,~\cite{yao2023retroformer,wang2024re2llm} propose reflection-based methods where LLMs refine recommendations by analyzing prediction against user interaction (Fig.~\ref{fig:more:simple}). While effective, they face three key challenges:

\textbf{(1) Ambiguous separation of explicit features and implicit patterns within one single user's interaction history.}
Effectively modeling user preferences in SeqRec requires distinguishing between explicit and implicit signals within user interaction histories. Explicit features are reflected in directly observable item features, such as titles, while implicit patterns are inferred from latent factors in user behavior, such as brand loyalty.
Current reflection-based recommendation methods~\cite{yao2023retroformer,wang2024re2llm} predominantly rely on explicit item features to model user interests, assuming that LLMs can effectively capture preference patterns from surface-level semantics alone. However, this reliance lacks analysis or reflection on the deeper, implicit connections that often drive user behavior. As illustrated in Fig.~\ref{fig:more:bc}, consider a user who has interacted with several electronic products and later purchases a T-shirt. If the recommender focuses solely on explicit item titles, it struggles to analyze the subtle connection between these interactions and fails to adapt to evolving user interests. In this case, the underlying implicit preference is the brand ``Apple'', reflecting a deeper pattern that explicit features alone cannot reveal.

\textbf{(2) Underutilized collaborative filtering (CF) signals across multiple user interaction histories.}
CF leverages behavioral similarities across users to detect shared interests (e.g., Apple product fans). Current approaches~\cite{yao2023retroformer,wang2024re2llm} neglect such signals, limiting their ability to exploit collective behavioral patterns for improved recommendations.

\textbf{(3) Suboptimal reflection update strategies.}
Existing reflection-based methods~\cite{wang2024re2llm,yao2023retroformer} mainly maintain a fixed reflection pool, lacking personalized dynamic maintenance based on user interactions. This inflexibility prevents them from adapting to users' dynamic preferences.

Facing these challenges, we propose \texttt{MoRE} (\underline{M}ixture \underline{o}f \underline{RE}flectors) to address the limitations through three coordinated innovations. \textbf{First}, on the offline side, we design specialized reflectors to model both \textit{intra-user patterns} and \textit{cross-user patterns}: 1) Two intra-user perspectives disentangle explicit features (e.g., product titles) and implicit behavioral patterns (e.g., brand loyalty) from a single user's interaction history; 2) A cross-user perspective captures collaborative filtering (CF) signals through item-centric analysis of user-item-user similarity relationships.
This decomposition directly resolves Challenges 1 (explicit/implicit ambiguity) and 2 (CF underutilization). 
\textbf{Second}, also on the offline side, our self-improving meta-reflector implements a lightweight two-phase update strategy: evaluating reflection quality via LLMREC performance gains (both presence/absence and old/new reflections), then iteratively generating improved reflections using top-performing candidates in an offline loop. This approach overcomes Challenge 3 (suboptimal updates) while maintaining computational efficiency. 
\textbf{Third}, on the online side, the meta-reflector dynamically selects optimal reflections per user through contextual bandit optimization~\cite{Li2010Contextual}. 
Experiments demonstrate \texttt{MoRE}'s superiority over traditional methods and LLMREC variants (prompt-based, fine-tuning-based, and reflection-based baselines) while maintaining low GPU memory and time costs.

In summary, our contributions are as follows:
\begin{itemize}[leftmargin=*]
    \item We first propose a dynamic LLM reflection framework is proposed for sequential recommendation to model and learn dynamic user preferences.
    \item \texttt{MoRE} incorporates three reflectors, each collecting reflections from explicit features, implicit patterns, and CF signals. It also integrates a meta-reflector, which updates the reflections with a refining-and-iteration strategy and selects the appropriate reflection for LLMREC using a contextual bandit algorithm.
    \item Extensive experiments on three real-world datasets demonstrate that \texttt{MoRE} outperforms state-of-the-art approaches in terms of recommendation performance with minimal GPU memory usage and training time overhead.
\end{itemize}

\section{Related Works}
\subsection{Sequential Recommendation}

SeqRec predicts users' next interests by modeling historical interactions chronologically, with the key challenge in capturing dynamic  preferences~\cite{shen2023hyperbandit,shen2024generating,zhang2024modeling, zhang2024reinforcing, zhang2025test,shi2024unisar,zhang2024saqrec,tang2025think}. Recent advances \citep{hidasi2015session,tang2018personalized,li2017neural} employ convolutional and recurrent neural networks, while \citet{li2022mlp4rec,li2023automlp,zhou2022filter} demonstrate MLP-based approaches achieve competitive performance. \citet{kang2018self,sun2019bert4rec,he2021locker,he2022query} utilize transformer-based models for item relevance modeling, achieving SOTA results.
Despite progress, existing methods remain limited in holistic preference modeling due to insufficient world knowledge and reasoning capabilities. This has spurred recent exploration of LLM-based approaches for sequential recommendation.

\begin{figure*}[t]
    \centering
\includegraphics[width=0.99\textwidth]{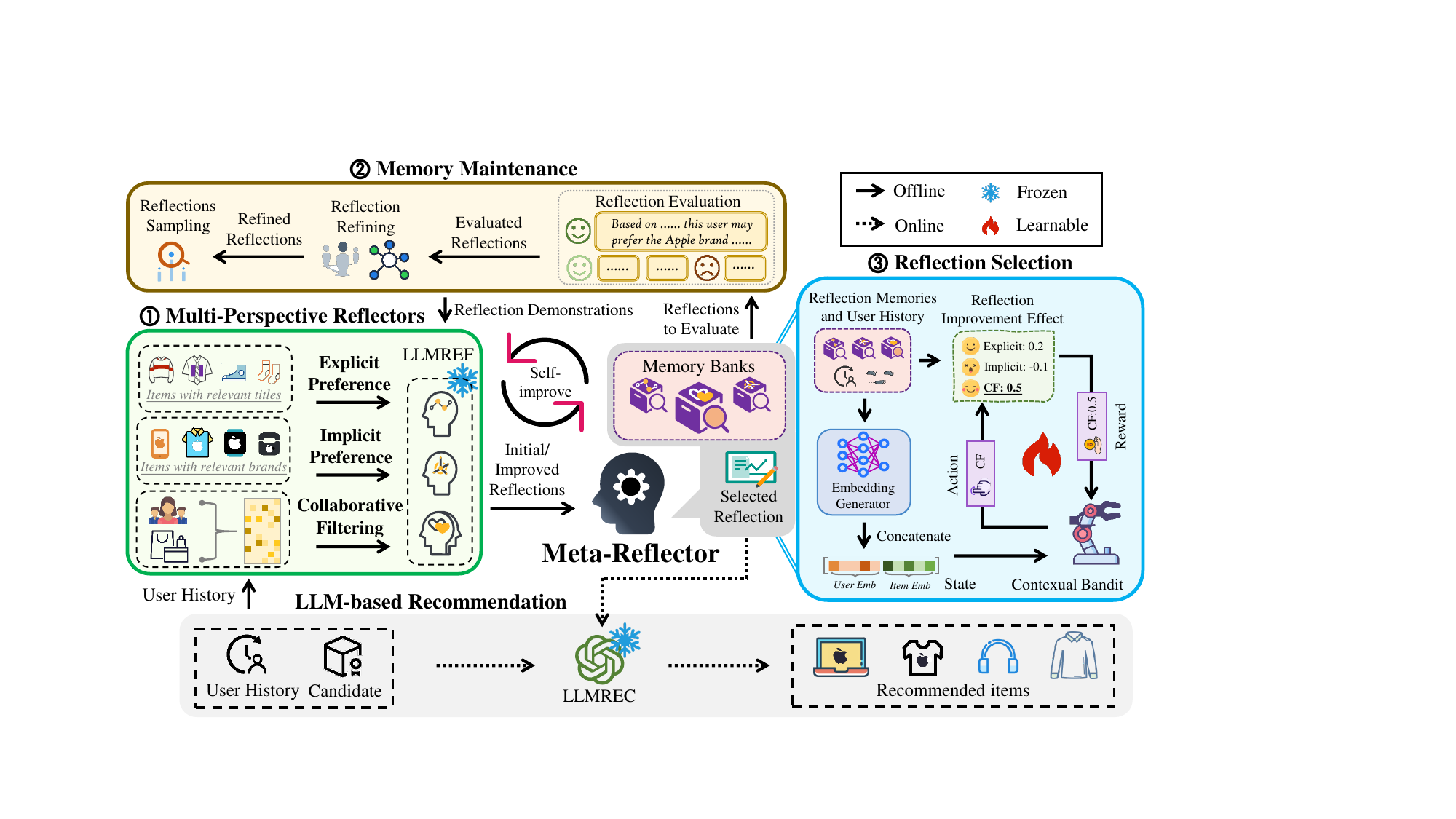}
    \caption{
    The overall architecture of \texttt{MoRE} framework. \texttt{MoRE} incorporates multi-perspective reflectors to generate reflections offline~\ding{172} and a meta-reflector to self-improve these reflections~\ding{173} and learn to select the most appropriate one~\ding{174}. During online recommendation~\ding{175}, the meta-reflector selects suitable reflections for the current user.
    }
    \label{fig:main}
\end{figure*}

\subsection{LLM-based Recommendation}

Current approaches to adapting LLMs for SeqRec predominantly fall into two categories: prompt-based methods and fine-tuning methods. Prompt-based methods involve manually constructing fixed prompts to guide LLMs in reasoning toward the desired items.
For example, LLM4RS~\cite{dai2023uncovering} 
enhances LLM's recommendation capabilities by customizing prompts to align with traditional ranking, 

while LLMRank~\cite{hou2024large} 
leverages LLMs for zero-shot ranking in recommendations through specialized prompting strategies. However, these fixed prompts rely heavily on human expertise and struggle to adapt to diverse users, often resulting in sub-optimal
performance~\cite{dong2022survey,wang2024re2llm}.
Fine-tuning methods{~\cite{geng2022recommendation,cui2022m6,zhang2023recommendation,bao2023tallrec,yue2023llamarec,liao2024llara,lin2023multifacet,shi2025unified}}, on the other hand, formulate SeqRec as a question answering task and fine-tune accordingly. To incorporate recommendation knowledge, they either add special tokens (e.g., Llara~\cite{liao2024llara}, LC-Rec~\cite{zheng2024adapting}) or insert embeddings as texts into the prompt (e.g., BinLLM~\cite{zhang2024text}).
Despite these efforts, a significant gap remains between LLM's pre-training and recommendation-oriented fine-tuning, which necessitates a substantial amount of tuning data to achieve proper alignment.
Additionally,{~\citep{hgsb,zhang2023collm,zhu2023collaborative,zheng2024adapting,zhang2024text,shi2025retrieval}} attempt to integrate collaborative filtering information into LLMs.
While effective, these methods lack an interpretation and analysis of the CF information, which prevents the integration of CF information with the powerful understanding and reasoning capabilities of LLMs, leading to potential suboptimal performance.

 Recently, \citet{wang2024re2llm} propose Re2LLM, a reflection-based approach that leverages LLMs to reflect on interaction history to enhance future recommendations. However, this approach maintains a static prompt pool without decoupled modeling of explicit and implicit preferences and incorporate CF signals.
 
\section{\texttt{MoRE}: The Proposed Framework}
\subsection{Problem Formulation}
In SeqRec, let $\mathcal{V} = \{v_1, v_2, ..., v_{|\mathcal{V}|}\}$ denotes the item set, $\mathcal{U} = \{u_1, u_2, ..., u_{|\mathcal{U}|}\}$ denotes the user set, and 
$\mathcal{S}_u^{t}=[v_u^1, v_u^2, ..., v_u^t]$ denotes the interaction sequence in chronological order for user $u\in\mathcal{U}$ up to time step $t$, where $v_u^{t} \in \mathcal{V}$ is the item that $u$ has interacted with at time step $t$. 
The goal of this task is to predict the item that $u$ will interact with at $t+1$ given $\mathcal{S}_u^{t}$:
\begin{equation}
\label{eq:rec}
    \hat{v}=\operatorname*{arg\,max}_{v\in \mathcal{V}}\Pr(v_{u}^{t+1} = v \mid \mathcal{S}_u^{t}).
\end{equation}
In this study, we focus on the LLM-based recommendation. Following~\cite{dai2023uncovering,hou2024large}, we adopt LLM to directly make the prediction in a ranking fashion as follows:
\begin{equation}
\label{eq:llm4rec}
\hat{\mathcal{O}}_u^{t+1}=\mathrm{LLM}_{\mathrm{REC}}(P_{\mathrm{REC}}(u, \mathcal{S}_u^{t}, \mathcal{C}_u^{t+1})), \hat{\mathcal{O}}_u^{t+1}\subseteq\mathcal{C}_u^{t+1},
\end{equation}
where $\mathcal{C}_u^{t+1}$ and $\hat{\mathcal{O}}_u^{t+1}$ respectively denote the candidate and predicted item list for $u$ at step $t+1$; all the parameters are concatenated in the text form; $\mathrm{LLM}_{\mathrm{REC}}$, the LLM for recommendation, is frozen; and $P_{\mathrm{REC}}$ denotes the manually crafted recommendation prompt.

More specifically, as illustrated in Fig.~\ref{fig:more:simple}, we focus on the two-stage reflection-based LLMREC. 

\textbf{In the offline reflection generation stage},  
given a user $u$, the interaction sequence $\mathcal{S}_u^{t-1}$ up to time step $t-1$ and the candidate list $\mathcal{C}_u^{t}$, we utilize LLM to make ranking prediction $\hat{\mathcal{O}}_u^t$ at step $t$:
\begin{equation}
\label{eq:ref-gen}
\hat{\mathcal{O}}_u^{t}=\mathrm{LLM}_{\mathrm{REC}}(P_{\mathrm{REC}}(u, \mathcal{S}_u^{t-1}, \mathcal{C}_u^{t}), \hat{\mathcal{O}}_u^{t}\subseteq\mathcal{C}_u^{t}.
\end{equation}

The LLM then reflects on the predicted list $\hat{\mathcal{O}}_u^t$ and the ground-truth item $v_u^t$ to infer $u$'s preferences:
\begin{equation}
\label{eq:ref-gen}
    Ref_{i,u} = \mathrm{LLM}_{\mathrm{REF}}(P_i(u,\mathcal{S}_u^{t-1},\mathcal{C}_u^{t}, \hat{\mathcal{O}}_u^{t}, v_u^{t})),
\end{equation}
where $\mathrm{LLM}_{\mathrm{REF}}$ denotes a frozen LLM responsible for generating reflections; the subscript $i\in\{EP, IP, CF\}$ for prompt $P_i$ and reflection $Ref_{i,u}$ corresponds to the ``Explicit Preference'', ``Implicit Preference'', and ``Collaborative Filtering'', which will be detailed in the next section.

\textbf{In the online recommendation stage}, the derived $Ref_{i,u}$ is incorporated into Eq.~\ref{eq:llm4rec} to enhance future recommendations:
\begin{equation}
\label{eq:ref-rec}
\hat{\mathcal{O}}_u^{t+1}=\mathrm{LLM}_{\mathrm{REC}}(P_{\mathrm{REC}}(u, \mathcal{S}_u^{t}, \mathcal{C}_u^{t+1}, Ref_{i,u})),\  \hat{\mathcal{O}}_u^{t+1}\subseteq\mathcal{C}_u^{t+1}.
\end{equation}

The details of recommendation prompt $P_{{REC}}$ are as follows:

\recboxsimple{\small{You are a recommender to recommend items for a specific user. The user 
interacted with items in the following order: $\langle S_u \rangle$. Reflections on the past 
recommendation attempt for this user (if any): $\langle Ref_{i,u} \rangle$. There are now $|C_u|$ 
candidate items: $\langle C_u \rangle$. Please consider the user's historical interaction sequence 
(and reflections), select appropriate items from the candidates, and rank 
them to recommend to the user. Think step by step. Your recommendations:}}

\subsection{Overall Framework}
We propose \texttt{MoRE}, a novel reflection-based framework, to model the dynamic preferences in LLM-based SeqRec. As shown in Fig.~\ref{fig:main}, \texttt{MoRE} consist of two key modules:

\textbf{Multi-Perspective Reflectors (offline reflection).} Motivated by the need to decouple explicit/implicit sequential features from user interaction histories and integrate user-item-user collaborative patterns, \texttt{MoRE} employs three perspective-aware reflection processes:

(1) Explicit/Implicit Decoupling: Two distinct reflections analyze a single user's history to disentangle explicit preferences (e.g., item titles and descriptions) and implicit preferences (e.g., attribute-level sequential patterns).

(2) CF-Aware Integration: A third reflection captures cross-user collaborative signals (e.g., rating trends and user similarity) through item-centric analysis.

As discussed in Sec.~\ref{sec:intro}, these derived reflections based on historical interactions are hints to enhance LLM's future recommendations. The details of this module will be introduced in Sec.~\ref{sec:multi:perspective}.

\textbf{Meta-Reflector (offline iteration and online selection).} 
This module has two main objectives. First, it maintains three memory banks for each user to collect reflections from the corresponding three reflectors and uses a refining-and-iteration strategy to self-improve the reflections and update the current reflection memories for each user. Second, it is responsible for deciding which of the three perspectives of reflection should be chosen to enhance the current recommendation at each step for the user. This decision-making process is formulated as a contextual bandit problem~\cite{Li2010Contextual}. These two parts will be elaborated in Sec.~\ref{sec:meta:main} and Sec.~\ref{sec:meta:selection}, respectively.

\subsection{Offline: Multi-Perspective Reflections}
\label{sec:multi:perspective}
As illustrated on the left of Fig.~\ref{fig:main}, \texttt{MoRE} aims at exploiting users' historical interaction to generate reflections from the perspectives of explicit preference, implicit preference, and collaborative signals. We achieve this goal through the following reflectors:

\textbf{Explicit Preference (EP) Reflector} 
captures user $u$'s explicit preferences by leveraging the LLM to reflect on the user’s interaction history $\mathcal{S}_u$, the candidate items $\mathcal{C}_u^t$, and the discrepancy between the predicted ranking $\hat{\mathcal{O}}_u^t$ generated from LLM-based recommender and the ground-truth $v_u^t$.

The reflection $Ref_{\mathrm{EP},u}$ from the perspective of explicit preference can be formulated as follows:
\begin{equation}
    \label{eq:EP}
    Ref_{EP,u} = \mathrm{LLM}_{\mathrm{REF}}(P_{EP}(u,\mathcal{S}_u^{t-1},\mathcal{C}_u^{t}, \hat{\mathcal{O}}_u^{t}, v_u^{t})),
\end{equation}
where $P_{{EP}}$ denotes the  prompt specifically designed to guide the LLM in reflecting explicit user preferences, defined as follows:

\epboxsimple{\small{You are a reflector for an LLM-based recommender system, understanding the explicit preferences embodied in the user's historical sequence and analyzing the areas for improvement in past recommendation attempts to provide reflections for the future. Explicit preferences are derived from an analysis of \textbf{recent tendencies reflected in the sequence of item titles and descriptions} within the user's history. You need to:
\newline
\textbf{1.} Analyze the history with associated text to identify explicit preferences.
\textbf{2.} Analyze the logic and rationale behind past recommendation attempts.
\textbf{3.} Examine potential shortcomings in the past and provide suggestions for improvement.
\newline
User's historical sequence with related description: $\langle REP_{EP}(S_u^{t-1}) \rangle$. Candidates: $\langle REP_{EP}(C_u^t) \rangle$. Past recommendation attempts: $\langle REP_{EP}(\hat{O}_u^t) \rangle$. User new interaction (if any): $\langle REP_{EP}(o_{u}^t) \rangle$. Historical reflection demonstrations (if any): $\langle DEMO \rangle$. Your reflection:}}

Please note that in EP reflector, item titles with descriptions are used to represent  $\mathcal{S}_u^{t-1},\mathcal{C}_u^{t}, \hat{\mathcal{O}}_u^{t},$ and $v_u^{t}$ through  $REP_{EP}()$.

\textbf{Implicit Preference (IP) Reflector} 
is designed to capture the preference embedded in the other item attributes (e.g., item brands).
Similar to the EP reflector, the IP reflector takes the interaction history $\mathcal{S}_u$ of user $u$, the candidate item list $\mathcal{C}_u$, the predicted ranking list $\hat{\mathcal{O}}_u$, and the ground-truth item $v_u^t$
as input, and produce the reflection through a frozen LLM:
\begin{equation}
    \label{eq:IP}
        Ref_{IP,u} = \mathrm{LLM}_{\mathrm{REF}}(P_{IP}(u,\mathcal{S}_u^{t-1},\mathcal{C}_u^{t}, \hat{\mathcal{O}}_u^{t}, v_u^{t})),
\end{equation}
where $P_{{IP}}$ denotes the prompt for guiding the
LLM in reflecting implicit user preferences, defined as follows:

\ipboxsimple{\small{You are a reflector for an LLM-based recommender system, understand the \textbf{implicit preferences} embodied in the user's historical sequence and analyze the areas for improvement in past recommendation attempts to provide reflections for future. Implicit preferences are reflected through the associations shown in subsequences of item attributes within the user's history, such as \textbf{subsequences of brands, styles, functions, features, etc.} Requirements:
\newline
\textbf{1.}  Analyze the logic and rationale behind past recommendation attempts.
\newline
\textbf{2.}  Focus on attribute subsequences within history to analyze potential associations and causality.
\newline
\textbf{3.}  Examine possible shortcomings in the past and provide suggestions.
\newline
Historical attribute seq: $\langle REP_{IP}(S_{u}^{t-1}) \rangle$. Candidates: $\langle REP_{IP}(C_u^t) \rangle$. Past recommendation attempts: $\langle REP_{IP}(\hat{O}_u^t) \rangle$. User new interaction (if any): $\langle REP_{IP}(o_{u}^t) \rangle$. Historical reflection demonstrations (if any): $\langle DEMO \rangle$. Your reflections:
}
}

In IP reflector, the attributes of items are used to represent $\mathcal{S}_u^{t-1},\mathcal{C}_u^{t}, \hat{\mathcal{O}}_u^{t},$ and $v_u^{t}$ through $REP_{IP}()$.

\textbf{Collaborative Filtering (CF) Reflector} aims to integrate patterns and similarities across user interaction sequences into LLM-based recommender systems. Unlike approaches that introduce new token embeddings into the vocabulary~\cite{zheng2024adapting,liao2024llara} or insert special sequences into prompts~\cite{zhang2024text}, we leverage a pre-trained collaborative filtering model $\mathcal{M}_u$ to incorporate CF ratings into the LLM reflection process.
Specifically, the CF reflector takes as input the target user $u$’s interaction history $\mathcal{S}_u^{t-1}$, the candidate item set $\mathcal{C}_u^t$, the predicted ranking $\hat{\mathcal{O}}_u^t$, and the ground-truth item $v_u^t$, and generates reflections using a frozen LLM:
\begin{equation}
    \label{eq:CF}
    Ref_{CF,u} = \mathrm{LLM}_{\mathrm{REF}}(P_{CF}(u,\mathcal{S}_u^{t-1},\mathcal{C}_u^{t}, \hat{\mathcal{O}}_u^{t}, v_u^{t})),
\end{equation}
where $P_{{CF}}$ is the prompt designed to guide the LLM in reflecting CF signals. The prompt structure is detailed as follows:

\cfboxsimple{\small{You are a reflector for an LLM-based recommender system, utilizing a Collaborative Filtering (CF) model to obtain items' \textbf{CF signals} and analyzing potential improvement in past recommendation attempts for future suggestions. The sequence of historical items' CF ratings reflects \textbf{patterns and similarities across user sequences}. Your step-by-step task:
\newline
\textbf{1.} Analyze historical CF ratings to identify preferences, trends, or interest shifts.
\newline
\textbf{2.} Evaluate past recommendation effectiveness by comparing suggested items with actual user interactions.
\newline
\textbf{3.} Analyze potential shortcomings in past recommendation attempts and provide suggestions for improvement.
\newline
Items presented with CF ratings: $\langle REP_{\mathcal{M}_u}(S_u^{t-1}, C_u^t, \hat{O}_u^t, o_u^t) \rangle$ Effective historical reflection demonstrations (if any): $\langle DEMO \rangle$. Your reflection:}}

In $P_{{CF}}$,
$\mathcal{S}_u^{t-1},\mathcal{C}_u^{t}, \hat{\mathcal{O}}_u^{t},$ and $v_u^{t}$ are represented by their corresponding CF ratings generated by $\mathcal{M}_u$ which is denoted as <$REP_{\mathcal{M}_u}$>.

For each user, we maintain three distinct memory banks, each dedicated to storing the generated reflections from one of the three perspectives (Explicit Preference, Implicit Preference, and Collaborative Filtering respectively). Each memory bank for a user stores a collection of past reflections for that user from the specific perspective.
These memory banks are guided by the meta-reflector to improve the reflections, as will be elaborated in the following section.

\subsection{Offline: Reflection Memory Maintenance}
\label{sec:meta:main}
Considering that LLMs may not always accurately generate reflection results, we need a mechanism to ensure the validity of reflections after the meta-reflector obtains them from the three reflectors. Therefore, we define a measure called the reflection improvement effect to assess the reflections and devise a refining-and-iteration strategy to update the reflections generated by the LLM.

\textbf{Reflection Improvement Effect} is defined as the difference in LLM recommendation performance on the validation set, with or without the reflections. In other words, if incorporating reflections improves the LLM's performance, it indicates that the reflection is effective for recommendation. The greater the performance improvement, the more effective the reflection. If the reflection improvement effect of a reflection exceeds a preset threshold $h$, we regard it as an effective one.

 Formally, we formulate the reflection improvement effect as:
\begin{equation}
\begin{aligned}
     \label{eq:refl:imp}
        Imp &= \mathrm{Metric}(\hat{\mathcal{O}}_{u}^\mathrm{REF}) - \mathrm{Metric}(\hat{\mathcal{O}}_{u}^\mathrm{REC}),\\
        \hat{\mathcal{O}}_{u}^\mathrm{REF} &=\mathrm{LLM}_{\mathrm{REC}}(P_{\mathrm{REC}}(u, \mathcal{S}_u^{t-1}, \mathcal{C}_u^{t}, Ref_{i,u})), \\
        \hat{\mathcal{O}}_{u}^\mathrm{REC} &=\mathrm{LLM}_{\mathrm{REC}}(P_{\mathrm{REC}}(u, \mathcal{S}_u^{t-1}, \mathcal{C}_u^{t})),
\end{aligned}
\end{equation}
where $Ref_{i,u}\in\{Ref_{\mathrm{EP},u},Ref_{\mathrm{IP},u},Ref_{\mathrm{CF},u}\}$, $\mathrm{Metric}$ denotes the recommendation metric, e.g., NDCG@10.  

\textbf{Refining.} 
As we filter out reflections that offer less or negative SeqRec improvement using the above measure, we further refine the selected reflections with three strategies:

(1) \textit{Global Level:} We select the reflections that bring the greatest average improvement for all users. In this approach, all users share the same reflections.

(2) \textit{Group Level}: We use a CF model to construct embeddings for all users and then cluster them using the K-means++ algorithm~\cite{bahmani2012scalable}. We then choose the reflections that provide the greatest average improvement for users within the same cluster, ensuring that users in the same cluster share the same reflections. 

(3) \textit{Individual Level:} We select the reflections that bring the greatest improvement for each user. This strategy leads to personalized reflections.

All of the three strategies are greedy, which may lead to local optima. To mitigate this issue, we propose using the improvement effect brought about by each reflection as the sampling probability. Reflections sampled from the filtered set serve as in-context learning demonstrations. These demonstrations are then used to task the three reflectors with generating improved reflections in subsequent iteration steps, enabling the LLM to self-improve its reflection quality.

\textbf{Iteration.} 
Since both reflection generation and refining are based on users' historical interactions, a major advantage of \texttt{MoRE} is that these two processes can be completed offline. 
As such, we can further use the sampled reflections as demonstrations and task the three reflectors with generating improved reflections in an in-context learning fashion~\cite{dong2022survey}. In this reflection generation and iteration loop, we enable the LLM to self-improve the reflections. We will further validate the impact of reflection iteration on LLM recommendation performance in the experiments.

\subsection{Online: Reflection Perspective Selection}
\label{sec:meta:selection}
The second objective of the meta-reflector is to model preference shifts in SeqRec. Given three memory banks for each user to collect effective reflections from three different perspectives, we require the meta-reflector to select one of these reflections for recommendation to achieve this goal.
We formulate the decision-making process as a Multi-Armed Contextual Bandit (MACB) problem~\cite{Li2010Contextual}, and adopt the Proximal Policy Optimization (PPO) algorithm~\cite{schulman2017proximal} to efficiently learn to select the reflections. Formally,
we define MACB as a tuple $(\mathcal{Z}, \mathcal{A}, \mathcal{R})$, where $\mathcal{Z}, \mathcal{A}, \mathcal{R}$ denote the state space, action (arm) space, and reward function, respectively.

(1) \textbf{state} $\bm{z} \in \mathcal{Z}$ represents recommendation context, constructed by concatenating CF-derived embeddings: 
$\bm{z} = [emb_u, emb_v]$ where $emb_u$ is the user embedding and $emb_v$ the average interaction embedding from $\mathcal{S}_u^{t-1}$.

(2) \textbf{Action} $\bm{a} \in\mathcal{A}$ is defined as the selection of a reflection from three memory banks. We represent $\bm{a}$ as a 3-dimensional vector, with each dimension corresponding to an arm, each arm representing a reflection memory.
    
(3) \textbf{Reward} is defined to assess the recommendation improvement brought by the reflections. 
    We reuse the measure in Eq.\eqref{eq:refl:imp} and represent the reward as $R(\bm{z, a})=Imp$. 
    Since the reflection has been assessed in Sec.~\ref{sec:multi:perspective} we do not need to call the LLM here. Additionally, due to the sampling discussed in Sec.~\ref{sec:meta:main}, $R(\bm{z, a})$ can be both positive and negative, ensuring training efficiency.
    
(4) \textbf{Replay Buffer} is designed to facilitate efficiency in policy optimization. We represent it as $D = (\bm{z}, \bm{a}, R(\bm{z},\bm{a}), \bm{z}')$ to store the tuples of observed state, action, reward, and next state. With the records in the replay buffer, we can refine successful policies and learn from erroneous trials. 
\label{sec:meta:ppo}

\begin{table}[!tbp]
\centering
\caption{Statistics of the 3 pre-processed datasets.}
\label{tab:dataset}
{\small\begin{tabular}{lcccc}
\toprule
\textbf{Dataset}         & \textbf{\#Users}   & \textbf{\#Items}   & \textbf{Avg. len}  & \textbf{Sparsity} \\ \midrule
\textbf{Arts}    & 55970           & 22612            & 8.80              & 99.96\%          \\
\textbf{Games}   & 55145           & 17287            & 9.01              & 99.95\%          \\
\textbf{Instruments} & 27404        & 10450            & 8.41              & 99.92\%          \\ \bottomrule
\end{tabular}}
\end{table}

\begin{table*}[!htbp]
\centering
\small
\setlength{\tabcolsep}{3pt}
\caption{
Recommendation performance comparison. All LLM-based methods utilize Llama-3 and ensure an identical candidate set for a fair comparison.
The best and the second-best performances are denoted in \textbf{bold} and \underline{underlined} fonts, respectively. ``N@K'' is short for ``NDCG@K''. The ``Imp.'' indicates the percentage improvement of \texttt{MoRE} over the best performances from baselines. ``-'' means that the baseline can only output a top-1 prediction and cannot rank or recall multiple items. $^\dagger$ denotes \texttt{MoRE} performs significantly better than baselines based on two-tailed paired t-test with Bonferroni correction ($p < 0.05$).}
\label{tab:exp:big}
\begin{tabular}{lccccccccccccccc}
\toprule
\multicolumn{1}{c|}{\multirow{2}{*}{Methods}} & \multicolumn{5}{c|}{Arts}                                                                                 & \multicolumn{5}{c|}{Games}                                                                                & \multicolumn{5}{c}{Instruments}                                                           \\ \cmidrule(l){2-16} 
\multicolumn{1}{c|}{}              & HR@1           & HR@5                 & HR@10                & N@5               & \multicolumn{1}{c|}{N@10}     & HR@1      & HR@5                 & HR@10                & N@5               & \multicolumn{1}{c|}{N@10}      & HR@1     & HR@5                 & HR@10                & N@5               & N@10              \\ \midrule
\multicolumn{1}{l|}{Caser}  & 0.103  & 0.2555                                            & 0.3603                                            & 0.1823                                            & \multicolumn{1}{c|}{0.2161}           & 0.0899                                   & 0.3024 & 0.4397 & 0.1997 & \multicolumn{1}{c|}{0.2440} & 0.1349   & 0.2931                                            & 0.3968                                            & 0.2193                                            & 0.2527                                            \\
\multicolumn{1}{l|}{FDSA}   & 0.1211   & 0.2536                                            & 0.3670                                             & 0.1875                                            & \multicolumn{1}{c|}{0.2241}       & 0.1196                                       & 0.3073 & 0.4466 & 0.2150  & \multicolumn{1}{c|}{0.2597} & 0.1403  & 0.2842                                            & 0.3905                                            & 0.2157                                            & 0.2497                                            \\
\multicolumn{1}{l|}{BERT4Rec}& 0.0896  & 0.2402                                            & 0.3613                                            & 0.1680                                             & \multicolumn{1}{c|}{0.2072}   & 0.0899                                           & 0.2796 & 0.4160  & 0.1861 & \multicolumn{1}{c|}{0.2299} & 0.1233  & 0.2618                                            & 0.3789                                            & 0.1954                                            & 0.2326                                            \\
\multicolumn{1}{l|}{GRU4Rec}  & 0.1001  & 0.2431                                            & 0.3708                                            & 0.1722                                            & \multicolumn{1}{c|}{0.2129}        & 0.0929                     & 0.3192 & 0.4644 & 0.2083 & \multicolumn{1}{c|}{0.2554}  & 0.1340     & 0.3021                                            & 0.4021                                            & 0.2201                                            & 0.2520                                             \\

\multicolumn{1}{l|}{SASRec}     & 0.1850          & 0.3570               & 0.4280               & 0.2788               & \multicolumn{1}{c|}{0.3016}   & 0.1900      & 0.2915               & 0.4348               & 0.1898               & \multicolumn{1}{c|}{0.2369}    & 0.1832    & 0.3327               & 0.3701               & 0.2594               & 0.2809               \\ \midrule
\multicolumn{1}{l|}{LLM4RS}        & 0.0543        & 0.1087               & 0.1277               & 0.0826               & \multicolumn{1}{c|}{0.0887}   & 0.0662      & 0.1976               & 0.2233               & 0.1338               & \multicolumn{1}{c|}{0.1421}    & 0.0330   & 0.0900               & 0.1188               & 0.0624               & 0.0717               \\
\multicolumn{1}{l|}{LLMRank(CoT)}        & 0.1725     & 0.3337			              & 0.4061              & 0.2595         & \multicolumn{1}{c|}{0.2828}   & 0.1611   & 0.3508	  & 0.4496      & 0.2607    & \multicolumn{1}{c|}{0.2925}   & 0.1457   & 0.2904		        & 0.3709              & 0.2202	                   & 0.2461              \\ 
\multicolumn{1}{l|}{LC-Rec}      & 0.2145      & 0.3768               & 0.4309               & 0.3006               & \multicolumn{1}{c|}{0.3181}    & \textbf{0.2599}    & 0.4140               & 0.4646               & \underline{0.3426}         & \multicolumn{1}{c|}{\underline{0.3593}}  & 0.1592    & 0.3296               & 0.3920               & 0.2486               & 0.2691      \\
\multicolumn{1}{l|}{Re2LLM}     & 0.2736      & \underline{0.4252}         & \underline{0.4938}         & \underline{0.3535}               & \multicolumn{1}{c|}{\underline{0.3757}}    & 0.2223    & \underline{0.4180}         & \underline{0.5198}         & 0.3231               & \multicolumn{1}{c|}{0.3559}    & \underline{0.2011}    & \underline{0.3521}         & \underline{0.4424}         & \underline{0.2787}         & \underline{0.3074}         \\
\midrule
\multicolumn{1}{l|}{\textbf{\texttt{MoRE}}}     & \textbf{0.2898 }        & \textbf{0.4376}      & \textbf{0.5043}      & \textbf{0.3705}      & \multicolumn{1}{c|}{\textbf{0.3922}}  & \underline{0.2569}  & \textbf{0.4773}      & \textbf{0.5731}      & \textbf{0.3647}      & \multicolumn{1}{c|}{\textbf{0.3957}} & \textbf{0.2163}  & \textbf{0.3824}      & \textbf{0.4674}      & \textbf{0.2981}      & \textbf{0.3251}      \\ \midrule
\multicolumn{1}{l|}{\textbf{Imp.}}    & \multicolumn{1}{c}{5.92\%$^\dagger$}     & \multicolumn{1}{c}{2.92\%$^\dagger$} & \multicolumn{1}{c}{2.13\%$^\dagger$} & \multicolumn{1}{c}{4.81\%$^\dagger$} & \multicolumn{1}{c|}{4.39\%$^\dagger$}    & \multicolumn{1}{c}{-1.15\%}              & \multicolumn{1}{c}{14.19\%$^\dagger$} & \multicolumn{1}{c}{10.25\%$^\dagger$} & \multicolumn{1}{c}{6.45\%$^\dagger$} & \multicolumn{1}{c|}{10.13\%$^\dagger$}       & \multicolumn{1}{c}{7.56\%$^\dagger$}       & \multicolumn{1}{c}{8.61\%$^\dagger$} & \multicolumn{1}{c}{5.65\%$^\dagger$} & \multicolumn{1}{c}{6.96\%$^\dagger$} & \multicolumn{1}{c}{5.76\%$^\dagger$} \\ \bottomrule
\end{tabular}
\end{table*}

\subsubsection{Training}
We apply the PPO algorithm~\cite{schulman2017proximal} with an Actor \& Critic network and a Clip objective function to efficiently learn  MACB.

(1) \textbf{Actor \& Critic Definition.} To model the selection of reflections and evaluate the value of the states, we implement a policy network parameterized by MLPs to define our selector’s Actor $\pi_{\bm{{\theta}}}$ and Critic $V_{\bm\psi}$, which maps the environmental space $\mathcal{Z}$ to the action space $\mathcal{A}$ and reward function $R$ respectively:
        \begin{equation}
        \label{eq:ac}
        \begin{aligned}
        \mathrm{Actor\ } \pi_{\bm\theta}(\bm{z}): \bm{a} &= \mathrm{softmax}(\bm{\theta}\cdot \bm{z}), \\
        \mathrm{Critic\ } V_{\bm\psi}(\bm{z}): r &= \bm{\psi}\cdot \bm{z},
        \end{aligned}
        \end{equation}
    where $\bm{\theta}$ and $\bm{\psi}$ are the parameters of Actor and Critic respectively. 
    
(2) \textbf{Clip Objective.} During the training process, the selector's Actor uses the $\epsilon$-greedy ($\epsilon = 0.1$) strategy to explore the environment, with probability $\epsilon$ to take a random action while with probability ($1-\epsilon$) to exploit the learned policy $\pi_{\bm{\theta}}$. We adopt the clip objective function for PPO training to maximize the reward selection as follows:
    \begin{equation}
    \label{eq:ppo:train}
    \begin{aligned}
        \mathcal{L} = \mathrm{min} \left(\frac{\pi_{\bm{\theta}}}{\pi_{\mathrm{old}}}A(\bm{z,a}), clip(\frac{\pi_{\bm{\theta}}}{\pi_{\mathrm{old}}}, 1-\delta,1 + \delta)A (\bm{z,a}) \right), 
    \end{aligned}
    \end{equation}
    where $A(\bm{z,a}) = \mathbb{E}(R(\bm{z} ,\bm{a})) - V_{\bm\psi}(\bm{z})$ is advantage function used to evaluate $\bm{z}$'s cumulative rewards $\mathbb{E}(R(\bm{z} ,\bm{a}))$ and current value $V_{\bm\psi}(\bm{z})$. $\delta$ is clip threshold.

\subsubsection{Inference} 
During inference, we select the reflection memories with the highest confidence from the Actor, exploit the reflection $Ref_{i,u}$ with the highest $Imp$ score in the memory, and make a recommendation according to Eq.\eqref{eq:llm4rec} using the prompt $P_{\mathrm{REC}}$.

\section{Experiment}

In this section, we conduct extensive experiments to answer the following research questions: 
\textbf{RQ1:} How does \texttt{MoRE} compare to various existing baselines in terms of recommendation performance? 
\textbf{RQ2:} How effective are the multi-perspective reflections and reflection perspective selection in \texttt{MoRE}? 
 \textbf{RQ3:} For reflection memory maintenance, how effective are the refining and iteration strategies? 
 \textbf{RQ4:} Does \texttt{MoRE} have an advantage in terms of training cost?

\begin{table}[!t]
\centering
\small
\caption{Comparison of HR@1 Performance with LLM baselines that inherently only provide top-1 prediction.}
\label{tab:exp:top-1}
\begin{tabular}{lccc}
\toprule
Methods   & Arts  & Games & Instruments \\ \midrule
TALLRec   & 0.2559      & 0.1975       & 0.1588            \\
BinLLM    & 0.2401      & 0.2347       & 0.1875            \\
LLaRA     & 0.2803 & 0.2263       & 0.1763            \\
A-LLMREC  & 0.2193      & 0.2342       & 0.1897   \\
MoRE      & \textbf{0.2898} & \textbf{0.2569} & \textbf{0.2163} \\ \bottomrule
\end{tabular}
\end{table}

\begin{table}[!t]
\setlength{\belowcaptionskip}{5pt}
\centering
\small
\caption{Ablation studies on Amazon Arts. ``Random'' indicates a random reflection selection from the three memory banks, while ``Greedy'' denotes that reflection selection is based solely on downstream SeqRec on the validation set. $Ref_\mathrm{all}$ denotes $Ref_\mathrm{EP}+Ref_\mathrm{IP}+Ref_\mathrm{CF}$. Note: ``+'' denotes concatenation, which may cause conflicting reflections leading to suboptimal performance. MoRE achieves the best results through user-personalized RL selection.
}
\label{tab:exp:ablation}
\begin{tabular}{l|ccccc}
\toprule
Methods           & HR@1     &HR@5 & HR@10  & N@5             & N@10            \\ \midrule
\textbf{\texttt{MoRE}} &\textbf{0.2898} &{\textbf{0.4376}} &{\textbf{0.5043}}                   & \textbf{0.3705} & \textbf{0.3922} \\ \midrule
- Random &{0.2736} &{0.4023} &{0.4881}              & 0.3421          & 0.3701          \\
- Greedy    &{0.2812} &{0.4156} &{0.4957}              & 0.3528          & 0.3786          \\ \midrule
- $Ref_\mathrm{EP}$ only   &{0.2850} &{0.4356} &{0.5033}        & 0.3670          & 0.3881          \\
- $Ref_\mathrm{IP}$ only   &{0.2850} &{0.4214} &{0.4881}        & 0.3571          & 0.3788          \\
- $Ref_\mathrm{CF}$ only     &{0.2850} &{0.4290} &{0.4871}      & 0.3572          & 0.3788          \\
- $Ref_\mathrm{EP}+Ref_\mathrm{IP}$   &{0.2583} &{0.4071} &{0.4700}      & 0.3443          & 0.3651          \\
- $Ref_\mathrm{EP}+Ref_\mathrm{CF}$   &{0.2707} &{0.3927} &{0.4623}      & 0.3311          & 0.3533          \\
- $Ref_\mathrm{IP}+Ref_\mathrm{CF}$     &{0.2402} &{0.3584} &{0.4156}    & 0.3033          & 0.3217          \\
- $Ref_\mathrm{all}$ &{0.2393} &{0.3594} &{0.4271}  & 0.3051          & 0.3268          \\ \bottomrule
\end{tabular}
\end{table}

\subsection{Experiment Settings}
\textbf{Dataset}. Following~\cite{zheng2024adapting}, we conduct our experiments on three subsets of the Amazon dataset~\cite{ni2019justifying}: ``Arts'', ``Video'', and ``Instruments''\footnote{Note that ``Arts'', ``Games'' and ``Instruments'' represent the abbreviations of ``Arts, Crafts and Sewing'', ``Video Games'' and ``Musical Instruments'' respectively in the Amazon Review dataset (2018).}. 
Each item in these subsets contains attributes that capture both explicit user preferences (e.g., item titles) and implicit preferences (e.g., attributes like brands and functional traits
). 
We follow~\cite{zheng2024adapting} to preprocess the data to achieve a fair comparison, with the statistical information shown in Tab.~\ref{tab:dataset}.

\textbf{Baselines}.
We adopt three types of SeqRec methods as baselines:
(1) Traditional Deep Learning Models: 
Caser~\cite{tang2018personalized} captures local and global patterns via CNN; 
GRU4Rec~\cite{hidasi2015session} models sequences with GRUs; 
SASRec~\cite{kang2018self} and BERT4Rec~\cite{sun2019bert4rec} employ uni-/bidirectional Transformers for next-item prediction; 
FDSA~\cite{zhang2019feature} models item-feature transitions via dual attention.
(2) Training-Free LLM Methods: 
LLM4RS~\cite{dai2023uncovering} aligns LLMs with IR ranking strategies; 
LLMRank~\cite{hou2024large} achieves zero-shot ranking via specialized prompting.
(3) Training-Required LLM Methods: 
LC-Rec~\cite{zheng2024adapting} expands vocabulary for collaborative filtering; 
TALLRec~\cite{bao2023tallrec} uses instruction tuning with text descriptions; 
BinLLM~\cite{zhang2024text} encodes user-item interactions as binary sequences; 
Re2LLM~\cite{wang2024re2llm} employs fixed reflection pools for suggestions; 
LLaRA~\cite{liao2024llara} combines user knowledge and behavior patterns via LoRA; 
A-LLMREC~\cite{hgsb} aligns CF embeddings with LLM vocabulary.

\textbf{Evaluation Metrics}. 
We adopt two widely used metrics, top-$k$ Hit Ratio (HR@$k$) and Normalized Discounted Cumulative Gain (NDCG@$k$), with $k\in\{5, 10\}$. We follow~\cite{zheng2024adapting, zhou2022filter} to employ the leave-one-out strategy for the obtaining of training, validation, and test data. Specifically, for each user behavior sequence, the last item is used as the test data, the penultimate item is used as the validation data, and the remaining interaction records are used for training. Following~\cite{hou2024large,wang2024re2llm,hgsb,liao2024llara}, we randomly sample $|\mathcal{C}_u|-1$ negative items to construct $|\mathcal{C}_u|$ candidates with 1 target item for each user.
Following ~\cite{wang2024re2llm}, $|\mathcal{C}_u|$ is set to 50 for all methods. 
All methods except those \textbf{inherently} limited to top-1 prediction (without ranking) must provide top-10 ranked items.

\textbf{Implementation Details}.
We use LLaMa-3-8B-Instruct~\cite{dubey2024llama} as the backbone for all LLM-based methods. Following~\cite{hou2024large,hgsb,bao2023tallrec,wang2024re2llm}, we respectively perform random sampling from three datasets for efficiency, and exclude users with overly long interaction sequences and then randomly we sample 1,000 users, with items remaining unchanged. For simplicity, we trained DMF\footnote{Please note that other CF models can be easily incorporated into \texttt{MoRE}.}~\cite{xue2017deep} with an embedding size of 64 to provide CF scores for reflection (Sec.~\ref{sec:multi:perspective}) and to assist with user clustering within refining (Sec.~\ref{sec:meta:main}). The number of clusters is set to 20. Threshold $h$ is set to 0.1.
We use RecBole~\cite{zhao2021recbole} to implement all traditional SeqRec baselines (e.g. SASRec) with the Adam optimizer and grid search\footnote{Grid search space was set as: \textit{embedding size} : \{32, 64, 128\}, \textit{learning rate} : \{$1e^{-2}, 1e^{-3},1e^{-4}$\} and \textit{batch size} : \{1024, 2048, 4096\}.}. 
For LLM baselines, we follow the origin settings and make proper adaptions to our scenario. For constrained generative retrieval baselines, we construct the constraints following the corresponding setting. For baselines that only provide top-1 prediction, we provide the same candidate set as MoRE and, following the original settings, do not require predictions beyond the top-1, thereby obtaining their best top-1 results under a lower task difficulty. 
For baselines limited to yes/no outputs, we adjust them, following LLaRA~\cite{liao2024llara}, to predict the next item. 
All experiments are run on a 4$\times$A800-PCIE-80GB GPUs server.

\begin{table*}[!t]
\small
\centering
\caption{
Training cost comparisons among LLM-based methods on four A800. ``Training time'' is measured in hours, shown as the mean and standard deviation (SD). ``VRAM'' denotes the Peak VRAM Usage, measured in GB. MoRE's RL framework minimizes computation cost through compact action space ($|\mathcal{A}|=3$, 3 perspectives).}
\label{tab:exp:cost}
\begin{tabular}{l|cc|cc|cc}
\toprule
\multicolumn{1}{c|}{\multirow{2}{*}{Methods}} & \multicolumn{2}{c|}{Arts}              & \multicolumn{2}{c|}{Games}            & \multicolumn{2}{c}{Instruments}       \\ \cmidrule(l){2-7} 
\multicolumn{1}{c|}{}                         & Training time                & VRAM        & Training time               & VRAM       & Training time                & VRAM       \\ \midrule
TALLRec (Fine-Tune)                                 &         26.228(±0.818)              & 46.98          &                 18.018(±0.290)      & 43.24         &   18.338(±0.487)                    & 42.95         \\
BinLLM (Fine-Tune)                                 &         10.209(±0.540)              & 124.67          &                 11.468(±0.328)      & 120.57         &   10.435(±0.451)                    & 115.48         \\
LLaRA (Fine-Tune)                                 &         9.013(±0.161)              & 77.92          &                 6.405(±0.281)      & 65.83         &   6.863(±0.187)                    & 74.36         \\
A-LLMREC (Fine-Tune)                                 &         56.975(±1.621)              & 217.74          &                  55.975(±1.058)      & 202.09        &    56.316(±1.992)                 & 231.63        \\
LC-Rec (Fine-Tune)                            & 20.382(±0.670)          & 206.37          & 21.361(±0.678)          & 201.35        & 13.720(±0.436)          & 199.93        \\
Re2LLM (RL, $|\mathcal{A}|=20$)                                   & 4.007(±0.007)          & 2.77          & 3.894(±0.005)          & 1.99          & 4.152(±0.008)          & 1.13          \\ \midrule
\textbf{\texttt{MoRE} (RL,  $|\mathcal{A}|=3$)}                            & \textbf{3.982(±0.001)} & \textbf{2.77} & \textbf{3.854(±0.002)} & \textbf{1.99} & \textbf{4.093(±0.003)} & \textbf{1.13} \\ \bottomrule
\end{tabular}

\end{table*}

\begin{figure*}[!ht]
    \setlength{\belowcaptionskip}{-5pt}
    \begin{subfigure}[h]{0.24\textwidth}
        \centering
    \includegraphics[width=\textwidth]{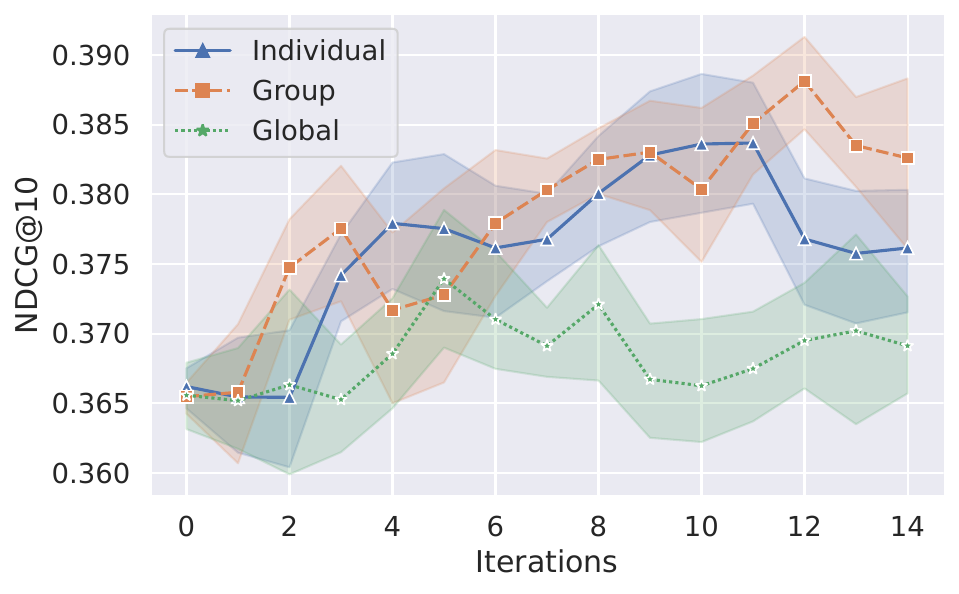}
        \caption{Arts}
        \label{fig:Arts:cluster:comp}
    \end{subfigure}
    \hfill
    \begin{subfigure}[h]{0.24\textwidth}
        \centering
        \includegraphics[width=\textwidth]{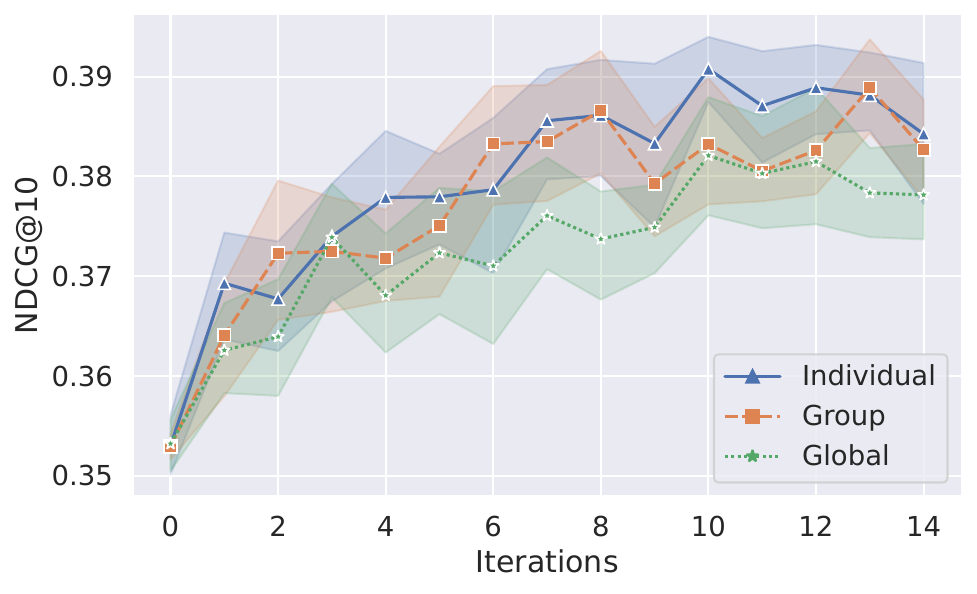}
        \caption{Games}
        \label{fig:Games:cluster:comp}
    \end{subfigure}
    \hfill
    \begin{subfigure}[h]{0.24\textwidth}
        \centering
        \includegraphics[width=\textwidth]{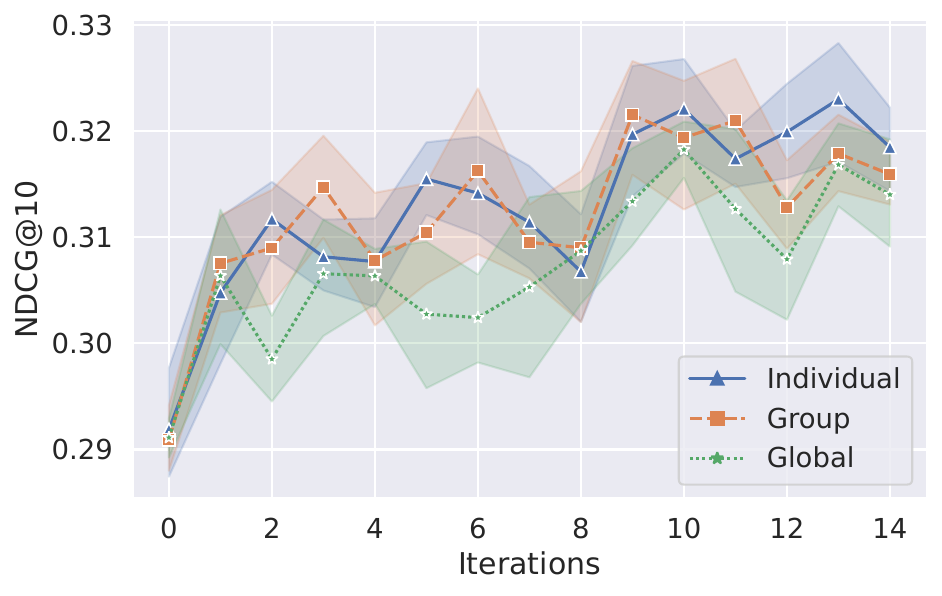}
        \caption{Instruments}
        \label{fig:Instruments:cluster:comp}
    \end{subfigure}
    \hfill
    \begin{subfigure}[h]{0.24\textwidth}
        \centering
        \includegraphics[width=\textwidth]{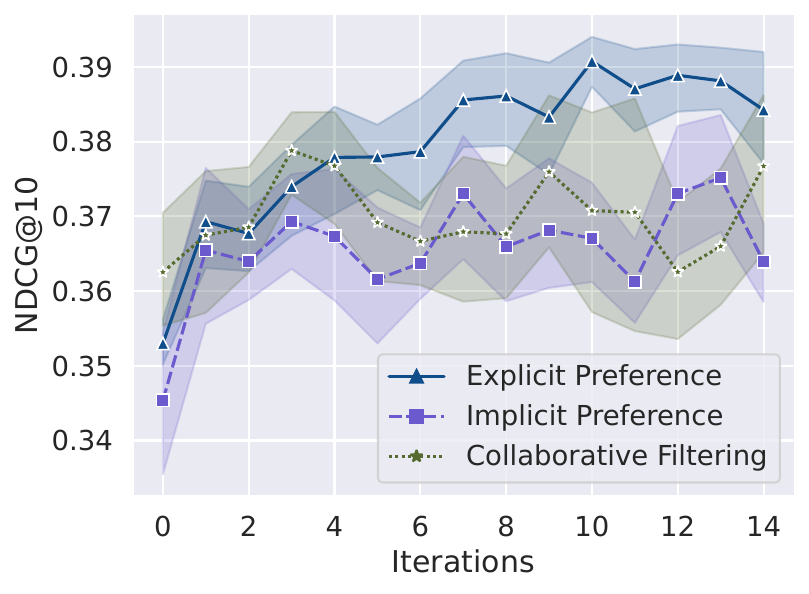}
        \caption{Iteration for 3 Perspectives}
        \label{fig:Instruments:cluster:comp}
    \end{subfigure}
    \caption{
    Impact of refining and iteration strategies on recommendation performance. 
    We repeated the experiment 5 times, with the mean value for each method represented by the line and the shaded area indicating the standard deviation. (a), (b), and (c) demonstrate the effects of iterations and different levels of refining on three datasets, while (d) shows the effects of the three perspectives with Games as an example.
    }
    \label{fig:cluster:comp}
\end{figure*}

\subsection{Recommendation Performance Comparison}
To answer \textbf{RQ1}, we compare the SeqRec performance of \texttt{MoRE} against baselines for 10 trials, with the average results presented in Tab.~\ref{tab:exp:big} and Tab.~\ref{tab:exp:top-1}. We have the following observations:
(1) \texttt{MoRE} outperforms all baselines across all metrics on all datasets, except for the suboptimal performance on the HR@1 on ``Games''. 
(2) In general, LLM-based methods demonstrate performance comparable to the traditional methods, which validates the potential of LLMs in SeqRec.  
(3) Notably, \texttt{MoRE} outperforms LLM-based methods that specialize in predicting one single item (HR@1 only), even when those methods sacrifice ranking capabilities\footnote{
Note that these methods are \textbf{inherently} limited to top-1 predictions. Therefore, for these methods, we retain only top-1 results for a fair comparison here.}.

In summary, \texttt{MoRE} exhibits superior performance in SeqRec, validating the efficacy of our approach.  We attribute this to our proposed LLM-based reflection framework, which leverages LLM's analytical and reasoning capabilities to capture users' explicit preferences and implicit behavioral patterns while seamlessly integrating CF knowledge in a clear and comprehensible manner, ultimately enhancing performance in sequence recommendation tasks.

\begin{figure}[!t]
\centering
\includegraphics[width=0.48\textwidth]{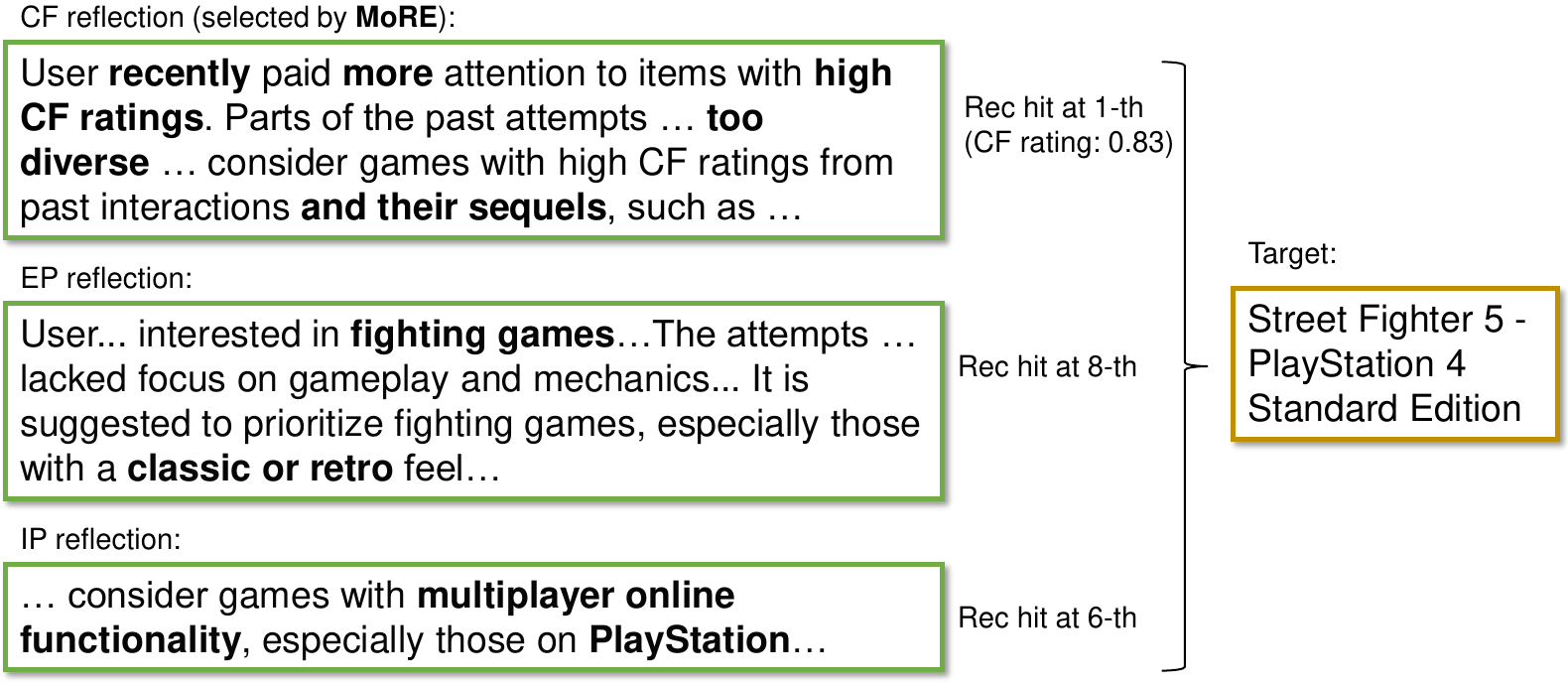}
\caption{
Example of how LLM reflections from explicit preference (EP), implicit preference (IP), and collaborative filtering (CF) perspectives improve recommendation.
}
\label{fig:case}
\end{figure}

\subsection{Ablation Study: Multi-Perspective Reflection and Selection}
\label{sec:exp:selection}
To validate the effectiveness of each perspective of reflection and answer \textbf{RQ2}, we conduct an ablation study on Amazon Arts. The results are presented in Tab.~\ref{tab:exp:ablation}.

We find that \texttt{MoRE}, equipped with the reflection perspective selection, outperforms alternatives, demonstrating the effectiveness of the multi-perspective reflections and the corresponding perspective selection. Specifically, \texttt{MoRE} outperforms the variants with not only the single reflection perspective but also the dual or triple perspectives simultaneously. The simultaneous use of multiple perspectives leads to a decrease in performance. This implies potential discrepancies among perspectives and no single perspective of reflection is universally optimal for all users, thereby confirming the necessity for selective choice among perspectives.

Moreover, \texttt{MoRE}'s selection strategy surpasses both the Random and Greedy selection approaches which respectively denote a random reflection selection from the three memory banks and the selection solely based on downstream recommendation performance.
This highlights the effectiveness of our contextual bandit modeling, enabling \texttt{MoRE} to choose the most suitable reflection for each user.

\subsection{Ablation Study: Refining and Iteration}
\label{exp:refine}
To answer \textbf{RQ3}, we test the impact of three refining strategies on SeqRec under different iteration rounds, taking reflections on explicit features as an example. 
From Fig.~\ref{fig:cluster:comp}, we find:

(1) \textbf{Iteration}: 
Overall, as the number of iterations increases, the recommendation performance gradually improves. This indicates that iteration can enhance the quality of reflections, thereby improving the recommendation.

(2) \textbf{Refining}: Refining at group and individual levels shows a significantly better trend over that at the global level during iterations, particularly on Arts (Fig.~\ref{fig:cluster:comp}(a)).  
We assume that the global-level reflections may introduce noise from other users, whereas group- and individual-level demonstrations more effectively elicit personalized reflective capabilities from the reflectors.

\subsection{Analysis of Computational Cost}

To address \textbf{RQ4}, we compare training costs (time and GPU memory) across LLM-based baselines using five repeated experiments, reported as Avg. $\pm$ SD\footnote{Average and standard deviation.} in Tab.~\ref{tab:exp:cost}.
We find: 1) Reflection-based methods (e.g., \texttt{MoRE}) incur lower costs than fine-tuning approaches by avoiding LLM parameter updates; 2) \texttt{MoRE} outperforms \texttt{Re2LLM} in efficiency due to its smaller action space.
\texttt{MoRE}'s Meta-Reflector selects from 3 reflection perspectives, while \texttt{Re2LLM} trains a PPO-based retrieval agent over a fixed base\footnote{The size is 20.}, requiring larger action-space exploration.
 In summary, \texttt{MoRE} achieves the lowest training costs (time and memory) among all baselines, demonstrating superior training efficiency.  

\subsection{Case Study}

We present a case study demonstrating reflections from EP, IP, and CF perspectives to enhance LLMREC (Fig.~\ref{fig:case}). The three perspectives improve performance as follows:
\textbf{EP Reflection} focuses on recent interaction trends (e.g., game genre preferences) to align with explicit preferences. This results in the target item being recommended at the 8-th position (e.g., \textit{Street Fighter 5}), effectively capturing short-term user intentions.
\textbf{IP Reflection} identifies patterns in historical attributes (e.g., multiplayer functionality, "PlayStation" brand) to adapt to implicit preferences. This moves the target to the 6-th position by analyzing deeper user behavioral patterns.
\textbf{CF Reflection} leverages collaborative filtering signals and emphasizes highly-rated items, reducing diversity while incorporating historical ratings and relevant sequels. This achieves the top-1 recommendation position. 
The \textbf{Meta-reflector} dynamically selects the optimal reflection (CF in this case), enabling \texttt{MoRE} to deliver optimal recommendations.

The reflection mechanism bridges the semantic gap between LLMs and CF: LLMs utilize simple score comparisons instead of interpreting embeddings. This approach cost-effectively equips LLMs with domain-specific knowledge (e.g., CF signals) while mitigating their adaptation challenges in specialized domains.

\section{Conclusion}
In this paper, we propose a mixture of reflectors framework, namely \texttt{MoRE}, for modeling and learning the dynamic user preferences in SeqRec.  
We first introduce three reflectors for each user to generate LLMs’ reflections from the perspectives of explicit user preferences, implicit user preferences, and collaborative signals. 
Building on these reflectors, we introduce a meta-reflector that evaluates and updates the generated reflections using a self-improving strategy. It selects the most appropriate perspective and corresponding reflections for each user's current recommendation using a contextual bandit algorithm. Furthermore, MORE's unique self-improving meta-reflector, through its rigorous refining and iteration strategies, explicitly addresses the critical challenge of ensuring the validity and quality of LLM-generated reflections, providing reliable hints for recommendation. This mechanism effectively bridges the semantic understanding of LLMs with established recommendation principles, enhancing both performance and interpretability.
Extensive experiments conducted on three benchmarks demonstrate that \texttt{MoRE} consistently outperforms traditional recommendation methods and LLM-based recommendation methods with minimal GPU memory usage and training time overhead. 

\newpage
\bibliographystyle{ACM-Reference-Format}
\bibliography{acmwww}

\end{document}